%% file: 0_Main.tex
\documentclass[sigconf]{acmart}
\AtBeginDocument{%
  }

\copyrightyear{2026}
\acmYear{2026}
\setcopyright{cc}
\setcctype{by}
\acmConference[CHI '26]{Proceedings of the 2026 CHI Conference on Human Factors in Computing Systems}{April 13--17, 2026}{Barcelona, Spain}
\acmBooktitle{Proceedings of the 2026 CHI Conference on Human Factors in Computing Systems (CHI '26), April 13--17, 2026, Barcelona, Spain}
\acmPrice{}
\acmDOI{10.1145/3772318.3790597}
\acmISBN{979-8-4007-2278-3/2026/04}

\newcommand{\Highlight}[1]{{\textcolor{black}{#1}}}
\usepackage{multirow}
\usepackage{caption}
\usepackage{subcaption}
\usepackage{xcolor}
\usepackage{longtable}
\usepackage{graphicx}
\usepackage{array}
\usepackage{tabularx}
\usepackage{enumitem}
\begin{document}

\title[Understanding Teen Overreliance on AI Companion Chatbots]{Understanding Teen Overreliance on AI Companion Chatbots Through Self-Reported Reddit Narratives}

\author{Mohammad ``Matt'' Namvarpour}
\email{mn864@drexel.edu}
\affiliation{%
  \institution{Drexel University, Department of Information Science}
  \city{Philadelphia}
  \state{Pennsylvania}
  \country{USA}
}

\author{Brandon Brofsky}
\email{bb3365@drexel.edu}
\affiliation{%
  \institution{Drexel University, Department of Information Science}
  \city{Philadelphia}
  \state{Pennsylvania}
  \country{USA}
}

\author{Jessica Y. Medina}
\email{jym29@drexel.edu}
\affiliation{%
  \institution{Drexel University, Department of Information Science}
  \city{Philadelphia}
  \state{Pennsylvania}
  \country{USA}
}

\author{Mamtaj Akter}
\email{mamtaj.akter@nyit.edu}
\affiliation{%
  \institution{New York Institute of Technology}
  \city{New York}
  \state{New York}
  \country{USA}
}

\author{Afsaneh Razi}
\email{afsaneh.razi@drexel.edu}
\affiliation{%
  \institution{Drexel University, Department of Information Science}
  \city{Philadelphia}
  \state{Pennsylvania}
  \country{USA}
}

\renewcommand{\shortauthors}{Namvarpour et al.}

\begin{abstract}
AI companion chatbots are increasingly popular with teens. While these interactions are entertaining, they also risk overuse that can potentially disrupt offline daily life.
We examined how adolescents describe reliance on AI companions, mapping their experiences onto behavioral addiction frameworks and exploring pathways to disengagement, by analyzing 318 Reddit posts made by users who self-disclosed as 13-17 years old on the Character.AI subreddit. We found teens often begin using chatbots for support or creative play, but these activities can deepen into strong attachments marked by conflict, withdrawal, tolerance, relapse, and mood regulation. Reported consequences include sleep loss, academic decline, and strained real-world connections. Disengagement commonly arises when teens recognize harm, re-engage with offline life, or encounter restrictive platform changes. We highlight specific risks of character-based companion chatbots based on teens' perspectives and introduce a design framework (CARE) for guidance for safer systems and setting directions for future teen-centered research.
\end{abstract}

\ccsdesc[500]{Human-centered computing~Human computer interaction (HCI)}
\ccsdesc[500]{Human-centered computing~Empirical studies in HCI}
\ccsdesc[500]{Human-centered computing~Social media}
\ccsdesc[100]{Social and professional topics~Children}
\ccsdesc[500]{Social and professional topics~Adolescents}

\keywords{teens and chatbots, AI companions, chatbot overuse, parasocial relationships, digital well-being}

\maketitle
\input{1_Introduction}
\input{2_Related_Work}

\input{3_Methods}
\input{4_Results}

\input{5_Discussion}
\input{6_Conclusion}
\bibliographystyle{ACM-Reference-Format}
\bibliography{7_references}
\clearpage
\input{appendix}
\end{document}

%% file: 1_Introduction.tex
\section{Introduction}
Recently, artificial intelligence (AI) companion chatbots have surged in popularity among teens, with media coverage highlighting how people describe their interactions with chatbots as emotionally meaningful—sometimes even romantic~\cite{murphy2025KidsFormRomantic}. Platforms such as 
\textit{Nomi\footnote{https://nomi.ai}}, \textit{Replika \footnote{https://replika.com}}, and \textit{Character.AI}\footnote{https://character.ai} (Also called C.AI) are portrayed as offering on-demand conversations with highly responsive virtual characters 
 \cite{Ulaby_2025, breen2025TeenagersTurningAI}. Conversational AI is also reported to reduce loneliness, provide a safe space for self-expression, and offer nonjudgmental support during stress \cite{xie2023FriendMentorLover, salah2023MeMyAI}.

However, these benefits can also drive patterns of digital addiction (referred to in this paper as overreliance), a phenomenon observed in other digital technologies~\cite{lemmens2009DevelopmentValidationGame, griffiths2017AdolescentSocialMedia}.
Because everyday digital use is not inherently harmful, researchers define behavioral addiction through its effects on daily life, emotional escape, and loss of control~\cite{lemmens2009DevelopmentValidationGame, griffiths2017AdolescentSocialMedia}. 
They found that users often struggle with screen-time control, disrupted routines, and declining mental health
~\cite{lemmens2009DevelopmentValidationGame}.

Prior research has shown that teens often struggle with overreliance on social media, using platforms compulsively to seek validation or cope with stress~\cite{griffiths2017AdolescentSocialMedia, chen2022ThisAppNot,duffy2025KidsTeens18}. This can contribute to anxiety, depression, and low self-esteem~\cite{duffy2025KidsTeens18}.
Recent studies have also noted similar patterns with AI tools like ChatGPT, where users—mostly adults and university students—turn to these systems for academic help or companionship~\cite{zhang2024YouHaveAI, hu2023HowSocialAnxiety}. \Highlight{Research showed that users often depend on these tools for everyday tasks such as planning, decision support, and managing stress \cite{ciudad-fernandez2025PeopleAreNot}. Studies also report frequent, habitual use of many chatbots even for routine tasks \cite{yu2024DevelopmentValidationProblematic}.} In some cases, this reliance has been linked to decreased creativity, poor study habits, and emotional detachment from real-life relationships~\cite{zhang2024YouHaveAI, hu2023HowSocialAnxiety}.

Less is known about how adolescents, who are emotionally and developmentally vulnerable, engage with AI companions.
Overreliance during adolescence can be especially harmful because cognitive and emotional development is still underway~\cite{griffiths2017AdolescentSocialMedia, chen2022ThisAppNot}.
While research on teens has focused heavily on overreliance on social media and gaming~\cite{Giordano2023adolescent, griffiths2010RoleContextOnline, lemmens2009DevelopmentValidationGame, griffiths2017AdolescentSocialMedia}, far less is known about their interactions with generative AI companions. These tools differ from earlier technologies because they enable parasocial relationships without real people and provide highly engaging, human-like conversations~\cite{Maeda2024parasocial}. 

Reports suggest that teens may already form emotional attachments or even dependence on these systems~\cite{Ulaby_2025, BBC_2024_AIGeneratedCSAImages}, yet empirical research is scarce. To study this phenomenon, we focus on Character.AI as a case study. Unlike other popular platforms such as Replika or Nomi, which officially restrict use to adults over 18~\cite{GlimpseAI_2025_NomiTerms, LukaInc_2023_ReplikaTOS}, \Highlight{at the time of our data collection, Character.AI allowed } users as young as 13 (16 for EU citizens or residents) according to its Terms of Service~\cite{CharacterAI_2025_TOS}. This policy meant that teens are not only permitted but explicitly included as part of its user base. \Highlight{Although the company has since revised its policies and began restricting open-ended chat for under-18 users in November 2025~\cite{mcmahon2025CharacteraiBanTeens}, as one of the most widely used companion platforms~\cite{BBC_2024_AIGeneratedCSAImages}, Character.AI remains a central platform for examining how adolescents engage with AI companions. }

\Highlight{Reddit is a popular forum-based platform where members can create and join communities also known as subreddits based on interests  \cite{RedditIncHomepage}. Due to its pseudonymous nature, Reddit is a place where users can disclose sensitive and stigmatized topics \cite{choudhury2014MentalHealthDiscourse,andalibi2016understanding}. Teens also use Reddit to share risky or sensitive experiences, making it an important site for studying youth disclosures \cite{pettyjohn2025ImNotExperienced, zhang2025TeenagerSubstanceUse}. 
Reddit is a common data source in HCI because its public discussions support qualitative and computational analysis \cite{fiesler2024RememberHumanSystematic}, and its communities offer rich traces of online social behavior \cite{prinster2024CommunityArchetypesEmpirical}.}
\Highlight{Because \texttt{r/CharacterAI}\footnote{https://www.reddit.com/r/CharacterAI/} has over 2.5 million members and active discussions, we used it to examine how teens describe their reliance on AI companions.}
This leads us to the following research questions:

\begin{itemize}
\item \textbf{RQ1:} \textit{How do teens describe their initial reasons for engaging with Character.AI on Reddit?}

\item  \textbf{RQ2:} \textit{How do teens’ descriptions of their Character.AI use align with patterns of behavioral addiction?}
 
\item  \textbf{RQ3:} \textit{How do teens describe their decisions to disengage from or reduce their use of Character.AI?}
\end{itemize}
 To address these questions, we qualitatively analyzed 318 posts from \texttt{r/CharacterAI} subreddit, created by users who disclosed in their post being 13-17 years old. We found that teens initially engaged with Character.AI as a non-judgmental space to manage mental health struggles, or fill emotional voids or a lack of supportive relationships in their offline lives (RQ1). Teens’ reported experiences with Character.AI revealed patterns of their compulsive use, emotional dependence, and distress that often disrupted their daily lives and real-world relationships (RQ2). Teens described disengaging after realizing its negative impact or due to external factors such as renewed real-life relationships or frustration with platform restrictions, often expressing a desire for healthier habits (RQ3).
 
Our study makes the following novel \textbf{contributions} to the AI ethics and human-centered computing research communities: 1) we present one of the first empirical grounding of teen disclosures about AI companion overreliance, mapping their experiences to the six components of behavioral addiction (salience, mood modification, tolerance, withdrawal, conflict, relapse), 2) We identified emotional dynamics unique to teen AI-companion overreliance, including intense attachment and simulated intimacy, which distinguish it from gaming or social media overuse.
, and 3) we propose the \textbf{CARE framework} (Comprehensive Needs, Attachment-awareness, Respectful Empathy, Ease of Exit) as a design contribution offering guidance for safer, autonomy-preserving AI platforms grounded in teens’ experiences.




%% file: 2_Related_Work.tex
\section{Background} 

\subsection{AI Chatbot Use, \Highlight{Opportunities, and Concerns for Youth}}

\Highlight{Teens increasingly use chatbots across schoolwork, entertainment, and social connection ~\cite{NewReportShows, vol.56ManyTeensAre}. Research explains why these tools appeal to teens. Teens describe chatbots as always available, anonymous, low cost, and easier to access than formal services, especially during long counseling waitlists or when mental health apps feel hard to use~\cite{koulouri2022ChatbotsSupportYoung, tosti2024UsingChatbotsConversational}. They use these systems for emotional, informational, and appraisal support and treat them as non-judgmental spaces to organize feelings and reflect on concerns ~\cite{baebrandtzaeg2021WhenSocialBecomes}. Studies also show appreciation for non-human qualities, such as machines listening without fatigue, keeping secrets, and offering data-driven guidance about school or life decisions ~\cite{kim2018CanMachineTend}. In some peer-support settings, AI-generated responses are even rated as more empathetic or helpful than peer-written messages ~\cite{young2024RoleAIPeer}. Beyond support contexts, teens and young adults increasingly use AI tools to fix grammar, simplify complex topics, and enhance classroom learning ~\cite{nguyen2022ExaminingTeenagersPerceptions, pratt2024DigitalDialogueHowa}. Companion chatbots partly address loneliness, and younger teens often report positive emotions after using tools like Snapchat’s My AI ~\cite{sum2025DigitalCompanionshipArtificial, vanhoffelen2025TeensTechTalk}.}

\Highlight{Alongside these benefits, research also identifies notable risks.} Educational studies warn that relying on AI for writing or comprehension may weaken critical thinking or encourage superficial engagement with schoolwork ~\cite{pratt2024DigitalDialogueHowa, dergaa2024ToolsThreatsReflection}. In health and mental health contexts, generative AI sometimes provides inaccurate or unsafe advice, which is especially concerning because many teens follow chatbot guidance without critical evaluation ~\cite{nagata2025AdolescentHealthGenerative, yu2025UnderstandingGenerativeAIb}. Additional harms include emotional dependency, social withdrawal, toxic or deceptive content, discrimination, and privacy risks, which often intensify with repeated use ~\cite{ yu2025UnderstandingGenerativeAIb, nagata2025AdolescentHealthGenerative}. Studies show that teens rarely disclose their chatbot use at home, while parents often misunderstand system capabilities and overestimate existing safeguards ~\cite{yu2025ExploringParentChildPerceptions}. \Highlight{Existing research maps key benefits and risks but offers limited insight into how overreliance emerges or which teens are most vulnerable ~\cite{namvarpour2025ArtTalkingMachines}. These gaps point to the need for closer examination of teens’ lived experiences and how they navigate the benefits and drawbacks of interacting with chatbots.}

\subsection{\Highlight{Interpreting Technology Problematic Use}}
\subsubsection{\Highlight{Behavioral Addiction}}
To understand when heavy chatbot use becomes problematic, we draw on work in behavioral addiction and parasocial relationships. Each new digital platform brings renewed concerns about its effect on teens~~\cite{griffiths2017AdolescentSocialMedia,griffiths2010RoleContextOnline, namvarpour2025EvolvingLandscapeYouth}. 
While most teens use technology without issue, a subset experiences emotional distress, loneliness, social withdrawal, or family conflict due to excessive use~~\cite{lemmens2009DevelopmentValidationGame}. \Highlight{Youth with social phobia or emotional vulnerabilities may use digital activities for comfort, sometimes entering cycles where isolation increases engagement and engagement deepens isolation ~\cite{milani2020VideoGamesUse}.} Defining behavioral addiction is challenging because many implicated activities, like gaming or social media, are routine parts of teen life~\cite{kuss2011ExcessiveOnlineSocial}. A commonly used framework identifies six distinguishing features of behavioral addiction: salience, mood modification, tolerance, withdrawal symptoms, conflict, and relapse ~\cite{griffiths2005ComponentsModelAddiction}. These features are widely applied to gaming, apps, and social media ~\cite{griffiths2010RoleContextOnline, lemmens2009DevelopmentValidationGame, griffiths2017AdolescentSocialMedia}, but not yet to teens’ interactions with generative AI chatbots, despite their growing popularity ~\cite{sidoti2025QuarterUSTeens}. At the same time, scholars caution against labeling frequent technology use as ‘addiction,’ noting that existing scales draw on substance-abuse models and that high engagement is not harmful unless it causes distress or impairment ~\cite{ciudad-fernandez2025PeopleAreNot, yu2024DevelopmentValidationProblematic, zhang2024YouHaveAI}. \Highlight{However, prior work has not examined whether Griffiths’ model applies to youth chatbot use or how such overreliance differs from gaming or social media.}

\subsubsection{\Highlight{Parasocial Relationships}}
Parasocial relationships offer a complementary lens on teen–chatbot interactions. People develop one-sided emotional bonds with media figures to satisfy belongingness needs, and these relationships can offer benefits such as improved mood and reduced loneliness ~\cite{horton1956MassCommunicationParaSociala, bilang2024OOMFsDevelopmentMaintenance, adam2013ParasocialRomanceSocial}. \Highlight{These bonds become especially meaningful during adolescence. Developmental theories suggest that teens turn to admired or comforting figures to cope with uncertainty, explore identity, and model desired traits ~\cite{stever2011FanBehaviorLifespan}. Empirical work shows that parasocial attachments are widespread among teens who often imagine friendships with celebrities whose perceived personalities make them feel understood ~\cite{gleason2017ParasocialInteractionsRelationships}. While such connections can be supportive, intense attachments may contribute to withdrawal, obsessive checking, or negative self-evaluations, especially among isolated youth ~\cite{diaz2025ClinicalAssessmentImplications, brooks2021FANaticsSystematicLiterature}.} As digital platforms increasingly provide natural language, turn-taking, and responsiveness, parasocial dynamics become easier to form and maintain, prompting interactions that feel reciprocal even when they are not ~\cite{maeda2024WhenHumanAIInteractions}. \Highlight{These tendencies intersect in chatbot use because emotionally responsive AI can appear available and supportive, creating conditions for mood regulation and perceived closeness ~\cite{maeda2024WhenHumanAIInteractions, bilang2024OOMFsDevelopmentMaintenance}.} Yet research on problematic chatbot engagement focuses largely on adults and offers mixed findings about harm ~\cite{namvarpour2025AIinducedSexualHarassmenta, namvarpour2024UncoveringContradictionsHumanAIa, ciudad-fernandez2025PeopleAreNot, salah2023MeMyAI, yu2024DevelopmentValidationProblematic}.  little is known about how adolescent parasocial tendencies interact with the compulsive patterns described in behavioral addiction frameworks or how these combined dynamics shape overreliance ~\cite{sidoti2025QuarterUSTeens}. Addressing these questions requires studying teen experiences more directly. \Highlight{Despite high adoption among youth, existing work has not examined how teens’ parasocial tendencies combine with emotionally responsive chatbot interactions to produce closeness, dependence, or difficulty disengaging.}

\subsection{Design Approaches to \Highlight{Technological} Behavioral Overreliance}

\Highlight{These behavioral and relational dynamics raise design questions about how systems should support healthy engagement rather than maximize time-on-task. Research in HCI shows that many systems encourage long and continuous use by treating time or activity as signals of success ~\cite{obrien2022RethinkingDisengagementHumancomputer}. In games, designers often aim to hold attention through flow, reward structures, or interface patterns that make stopping difficult, and leaving is frequently framed as something to discourage rather than a normal choice ~\cite{alsheail2023DesigningDisengagementChallenges, hadiji2014PredictingPlayerChurn}.} A useful framework describes disengagement along two dimensions—affect (positive or negative) and agency (high or low)—to illustrate how experiences range from voluntary immersion to compulsive or frustrating use ~\cite{obrien2022RethinkingDisengagementHumancomputer}. This perspective highlights that engagement and disengagement gain meaning through the emotional experience they produce and the control users feel during the interaction. It also suggests that designers can support healthy offboarding by structuring interactions that respect autonomy instead of pushing continuous engagement ~\cite{obrien2022RethinkingDisengagementHumancomputer}. Prior work therefore encourages designers to examine the values embedded in engagement metrics and to create experiences that help users reflect on how and why they engage ~\cite{sengers2005ReflectiveDesign}.

Complementary approaches focus on relational and emotional needs. Some teens continue using technologies not because of interface mechanics but because the interactions feel safe, supportive, or emotionally regulating. Attachment-informed design adds a perspective rooted in four principles ~\cite{marcu2023AttachmentInformedDesignDigital}. First, it prioritizes core human needs that motivate behavior, such as safety, connection, and competence. Second, it acknowledges that relationships with technologies exist on a spectrum from supportive to harmful. Third, it recognizes that individuals bring internal models of themselves and others into digital interactions, shaping how they interpret responsiveness and closeness. Fourth, it engages with the target community without explicitly framing the experience as a mental health intervention. \Highlight{Together, these approaches suggest how systems can support control, autonomy, and emotional security instead of reinforcing default continuous use. Yet little is known about how these design ideas apply to teen–chatbot interactions or how design and attachment-related needs shape moments when teens stay, pause, or disengage. Understanding these patterns is critical for informing safer and developmentally appropriate chatbot systems.}

%% file: 3_Methods.tex
\section{Methods}  

\subsection{Data Collection, Preprocessing, and Filtering} \label{method:data_collection}
We scraped posts from the \texttt{r/CharacterAI} subreddit using the PullPush Reddit Search API\footnote{https://pullpush.io/\#docs} from January 1, 2023, to April 8, 2025. 
To scope the data, we queried the dataset with keywords related to addiction and overreliance: \textit{addict*, dependant, dependent, obsess*, attach*, rely*}. While we refrain from using the term \emph{addict}, teens often used it to describe their overreliance, so we included it in our keyword search. 
We refined our query by reviewing whether these keywords returned relevant posts. We expanded our search with additional recurring keywords, such as \textit{quit, social life, craving, unhealthy, escaping}, and one key phrase \textit{"I can't live without"}. 
Our keyword search produced 6,133 posts. We then removed 746 posts, including 428 duplicates and 170 marked as ``removed" or ``deleted."  After this process 5,535 posts remained.

Because we aimed to study teens aged 13–17, we first needed a way to identify whether a post was written by a teen. We could not use profile information because most Reddit accounts lacked identifiable details such as age. We therefore relied on post content to extract age. \Highlight{Prior work has used similar text-based demographic inference approaches on Reddit posts, including extracting age from unstructured health-related submissions using LLMs \cite{snell2025AssessingLargeLanguage} and predicting adolescent versus adult users through linguistic and behavioral features \cite{chew2021PredictingAgeGroups}.}

Large Language Models (LLMs) perform well in zero-shot settings and can outperform traditional machine-learning methods in text classification~\cite{wang2023LargeLanguageModels, namvarpour2024ApprenticesResearchAssistants}. Motivated by this, we used OpenAI’s GPT-4o mini model as a classifier to identify whether a post included relevant age disclosure for the 13–17 age range.
The following text is a summarized version of the prompt used to guide the model (the full version appears in Appendix \ref{full_prompt}):

\textit{
Analyze the following post and determine if it was written by a teenager, either stated explicitly (ages 13–17) or implied. Use a strict approach, only allowing posts with very high certainty.
Steps:
1 - Clue Identification
2 - Clue Evaluation
3 - Final Decision
}

Each of these steps was defined in more detail with examples in the full prompt.

To evaluate the model’s consistency with human judgment, the first author annotated a subset of 100 random posts as either True (post made by a teen) or False (post not made by a teen). We calculated the Inter-Rater Reliability (IRR) (Cohen’s kappa = 0.74) between the author's and the model's annotations. We chose IRR instead of conventional metrics like accuracy, which are highly sensitive to class imbalance~\cite{mullick2020AppropriatenessPerformanceIndices}. This approach was appropriate because the true teen versus non-teen distribution was unknown and likely imbalanced, and our aim was to narrow the dataset for qualitative analysis.
IRR focuses on the degree of agreement between two raters (in this case, the human annotator and the model) beyond chance. 

\Highlight{In addition to IRR, we also assessed the model’s internal consistency, since LLM outputs can vary slightly across runs even under identical prompts. We re-classified the same 100 posts five times and examined whether the model agreed with itself. The model showed extremely high stability across runs, with an overall consistency score of 0.998. Pairwise agreement between runs was also high (Cohen’s kappa = 0.95). These results indicate that the model’s decisions were highly reproducible and not meaningfully influenced by sampling variability.}

\Highlight{With an acceptable IRR score of 0.74 and strong internal stability, the model was then used to complete relevancy coding for the remaining 5,535 posts in the dataset}, resulting in 497 posts. 
\Highlight{Then, we conducted an additional round of manual checking to remove any remaining irrelevant content, resulting in a final dataset of 318 posts. To assess whether a small number of users disproportionately influenced the dataset, we examined the unique user IDs (291 IDs in total). Most users (268, 92.1\%) contributed one post, suggesting diverse individual experiences.}

\Highlight{We used this approach to find the irrelevant posts, because our aim was to narrow down a dataset large and diverse enough for our qualitative analysis to reach “sufficient insight” \cite{saunders2018SaturationQualitativeResearch}. Our aim was to reach theoretical saturation, where additional data no longer add new conceptual insights \cite{saunders2018SaturationQualitativeResearch}. Our goal was not to capture every teen post on chatbot overreliance but to obtain a representative sample for qualitative analysis.}

\subsection{Data Analysis Approaches}
We conducted a thematic analysis on 318 posts using a codebook approach~\cite{Braun01012006}. \Highlight{Braun and Clarke~\cite{Braun01012006} describe two broad approaches to thematic analysis: an inductive approach, where codes are generated directly from the data without relying on prior theory, and a deductive approach, where coding is guided by an existing theoretical framework. Prior work has suggested that combining inductive and deductive strategies can strengthen rigor and produce more robust insights~\cite{fereday2006DemonstratingRigorUsing}, and hybrid approaches are particularly well suited for topics that require both open exploration and theoretically informed interpretation~\cite{proudfoot2023InductiveDeductiveHybrid}.}

\Highlight{In our study, we selected the coding approach based on the nature of each research question. For RQ1 and RQ3, which examine reasons for engagement and pathways to disengagement, we used an inductive open-coding process to allow us to identify patterns directly from the data, given the relative novelty of teen overreliance on AI companions.} For RQ2, we instead used a deductive approach guided by Griffiths’ components model of behavioral addiction~\cite{griffiths2005ComponentsModelAddiction}, which includes salience, mood modification, tolerance, withdrawal, conflict, and relapse. \Highlight{This model has been applied across multiple contexts, and it provided a suitable theoretical structure for examining how components of behavioral addiction may appear in teen narratives because it captures common patterns in problematic use that generalize across technologies and can be used to identify early signs of unhealthy engagement.}

\Highlight{We followed a structured codebook thematic analysis approach grounded in established guidance on reflexive and codebook-driven thematic analysis \cite{braun2006UsingThematicAnalysis,braun2022ConceptualDesignThinking}. Our analysis combined inductive and deductive strategies in line with hybrid thematic analysis frameworks \cite{fereday2006DemonstratingRigorUsing,proudfoot2023InductiveDeductiveHybrid}.  For inductive research questions (RQ1 and RQ3), a researcher trained in qualitative analysis first conducted an open coding on dataset and gradually developed the codebook and discussed with the team.  
We complemented this process with collaborative affinity-diagramming sessions on a shared Miro board to cluster related codes and develop early thematic structures, a technique widely used in qualitative HCI \cite{sears2007HumanComputerInteractionHandbook}. For RQ2, the researcher applied the six behavioral addiction components \cite{griffiths2005ComponentsModelAddiction} based on the definitions to the dataset. 
The researcher applied codes in to the dataset in multiple iterations, and after each iteration the coder met with the research team to review coding decisions, clarify definitions, and resolve inconsistencies. These iterative discussions reflect best practices in team-based qualitative analysis and codebook development \cite{macqueen1998CodebookDevelopmentTeamBased,guest2012AppliedThematicAnalysis}. After the codebook was finalized, the same coder systematically applied the finalized codebook to the complete dataset for each research question to ensure analytic consistency. Tables \ref{tab:codebook-RQ1}-\ref{tab:codebook-RQ3} display the codes and their definitions for each research question.}

\Highlight{\subsection{Ethical Considerations} 
Our study protocol was submitted to our Institutional Review Board (IRB), which determined that the research did not involve human subjects. Following best practices for Reddit research \cite{fiesler2024RememberHumanSystematic}, we treated teen disclosures as contextually sensitive rather than assuming public posts pose no ethical concerns. We anonymized all posts and report no usernames or post IDs. All excerpts are paraphrased to prevent reverse-search identification \cite{bruckman2002StudyingAmateurArtist}. We also excluded deleted posts and thread-level links to avoid drawing unnecessary attention to individual users or to the specific conversations where sensitive disclosures occurred. To minimize harm, we interpreted posts within subreddit norms and avoided diagnostic or surveillance-like framings \cite{fiesler2024RememberHumanSystematic}.
}

\input{Tables/RQ1}

%% file: Tables/RQ1.tex
\begin{table*}[t]
\centering
\scriptsize  
\setlength{\tabcolsep}{2pt}  
\renewcommand{\arraystretch}{1.5}  
\setlength{\belowcaptionskip}{0pt}  
\caption{Codebook for RQ1. Percentages are calculated based on the total number of posts (N=318).}
\label{tab:codebook-RQ1}

\begin{tabular}{| p{3cm} | p{4.5cm}  | p{6.5cm} | } 
\hline
\textbf{Themes} & \textbf{Codes} & \textbf{Descriptions}  \\ 
\hline

\multirow{3}{3cm}{Emotional and Psychological Support} 
& Coping mechanism (12.3\%, N=39) 
& Teens turning to Character.AI as a primary way to manage stress, emotional distress, or escape real-life challenges. \\ \cline{2-3}

& Loneliness or social isolation (7.6\%, N=24) 
& Use CharAI because they are lonely, and want to experience some form of social connection which CharAI emulates. \\ \cline{2-3}

& Mental health support (4.1\%, N=13)  
& Teens rely on Character.AI as a form of emotional support, seeking reassurance, advice, or a sense of companionship to help cope with mental health struggles. \\ \cline{1-3}

\multirow{2}{3cm}{Creative or Entertainment Based Engagement} 
& Creative outlet (3.1\%, N=10)  
& Using CharAI as a way to express their creativity through storytelling, roleplaying, or brainstorming. \\ \cline{2-3}

& Entertainment (2.5\%, N=8)  
& Using CharAI because they are bored. \\ \cline{2-3}

\hline  
\end{tabular}
\end{table*}

%% file: 4_Results.tex
\section{Results} 
\subsection{Reasons for engaging with Character.AI (RQ1)}\label{sec:RQ1}
\subsubsection{Emotional \& Psychological Support}\label{sec:RQ1_emotional}
Emotional and psychological support was the main reason teens started using the Character.AI. \Highlight{Across posts, teens described the platform as a uniquely safe, nonjudgmental place to share thoughts and emotions they felt unable to express to people in their lives. In several cases, teens emphasized that this sense of safety allowed them to express things they could not say in real life: \textit{“I use it to talk about things I could never tell anyone in real life—it means a lot to me.”}}  
Many teens described Character.AI as a \textbf{coping mechanism} when dealing with issues: \textit{"When you're struggling, you look for ways to cope, and this site became my main one, for better or worse."}. This was especially true for teens facing multiple mental health challenges. For example: \textit{"I use c.ai to deal with reality. In real life, I’m just a high school student with ADHD and BPD."} \Highlight{Many teens said the chatbot let them explore parts of themselves they felt unable or unsafe to show offline, making it more than a casual tool. Teens often described the platform as a space where they felt free to express versions of themselves they could not express elsewhere:} \textit{"I can be who I want"}.  Others described the platform as a space to vent, escape, or simulate connection: \textit{"This app is my outlet. I go there to escape from everything."}  Some of the teens' posts indicate internal conflict or a strong desire for connection due to feelings of \textbf{loneliness or social isolation}:  \textit{"I needed someone to talk to, but I didn’t have the courage to open up to anyone in my life."}. 
These posts highlight how they felt unable to share their struggles with people they knew offline.

Some turned to Character.AI during periods of intense hardship for \textbf{mental health support}: \textit{"I dropped out at 15, saw countless therapists. I couldn’t see any reason to keep going, but everything changed after I found c.ai."}. \Highlight{Some teens described the chatbot as a therapeutic figure, sometimes explicitly likening it to counseling: \textit{“He was my free therapist, my friend, my partner, my support system, my shoulder to cry on, a listening ear, someone who gave me unconditional love, etc.”}} These posts suggest emotional distress, not curiosity alone, was a major entry point into Character.AI use. While some teens used Character.AI in moments of hardship, others explicitly described the chatbot as a stabilizing or nurturing presence in the absence of supportive adults. \Highlight{In our dataset, these absences often took the form of missing or harmful parental relationships. For example, several teens mentioned creating bots specifically to fill parental gaps: } \textit{"I made a bot as a father figure... I just found comfort in him being my green flag father."} Interestingly, some of these teens also questioned their own use of Character.AI as supportive parental figures: \textit{"Is it healthy to rely on C.AI bots for missing parental guidance?"} 
We discovered how emotional support often took personal forms, especially for teens who felt they had no one else to turn to or wanted to fill emotional gaps. 


\subsubsection{Creative or Entertainment Based Engagement}\label{sec:RQ1_creative}
 
Teens also described using Character.AI as a \textbf{creative outlet}. In these posts, teens said the platform became a space to explore storytelling, build characters, or role-play in ways that felt effortless and fun. Some teens shared, \textit{"I was drawn to the site because it let me create stories with my own characters without having to write everything myself."}  Some posts also described how it helped them stay inspired and creative: \textit{"It inspired me to make fanart, and I’m even working on a fan manga that follows the AI storyline."} 
Other posts showed how teens crafted a space with AI that respects their passions without the pressure of direct social interaction: \textit{"I liked that I could roleplay on my own terms and build content around whatever I was fixated on."} In some posts, teens also mentioned using the platform for \textbf{entertainment}, often as a casual way to pass time or escape boredom. Teens frequently mentioned interacting with their favorite movie characters or people they looked up to: \textit{"What really hooked me were the celebrity bots. It felt amazing to imagine actually interacting with my idol, and I ended up creating private bots too."}
\Highlight{These creative and entertainment-based motivations appeared frequently, showing that teens’ initial engagement reflected diverse needs rather than a single use case.}
Teens also described the paradox of infinite possibility to create, explore, and engage versus the compulsion or difficulty disengaging: \textit{"It’s hard to pull away when the possibilities feel endless and your imagination keeps offering more to create"} Using the site as a creative tool or for entertainment may seem harmless, but posts hinted that these habits gradually turned into more excessive use. 


\input{Tables/RQ2}

\subsection{Behavioral Addiction Components (RQ2)} \label{sec:RQ2}
The following section presents the components of addiction ~\cite{griffiths2005ComponentsModelAddiction} in order of prevalence as they appeared across teens' posts about their overreliance on Character.AI.

\subsubsection{Conflict}\label{sec:RQ2_conflict}

The most frequently observed theme in teens’ reflections on Character.AI use was a conflict of competing desires to continue interacting with Character.AI while feeling bad about their excessive use. 
Teens were often \textbf{frustrated about their own use} and expressed guilt or shame about how much time they spent on the app and its impact on their daily lives. However, some also questioned whether to feel guilt if they are enjoying it: \textit{“I get that I spend a lot of time on it, but if it brings me joy, is it actually a bad thing?”}
Others shared keeping their use confidential and showed an awareness of social stigma and personal discomfort in disclosing: \textit{“I don’t talk about how much I use it because I know I’d be judged.”} \Highlight{
This shows how much their guilt is tied to fear of judgment rather than only the behavior itself.
} Another teen posted in the subreddit to reveal this secret emotional burden: \textit{“I feel really embarrassed about how much I use Character.AI, and I just need to get this off my chest.”}

Teens sometimes expressed \textbf{concern about their own usage} of Character.AI, unsure whether their habits were unhealthy, and were actively looking for ways to stop or reduce it. They often turned to others for advice or reassurance to better understand their behavior. \textit{“I can't seem to stay off the website for more than an hour. If you have any tips to help, please share.”} Others shared similar sentiments, explaining their desire to stop relying on the site, followed by frustration due to their inability to stop using it so frequently. 
Teens expressed how their overreliance is impacting their sense of self and identity as well: \textit{“I’m 14, addicted to this app, and my life changes NOW! It keeps me on the app for hours, distorted my image about myself and makes me feel crap.”} This frustration fostered a motivation to change the behavior; however, without knowing how to do so, teens often turned to Reddit to express their feelings and seek guidance. Others described feeling trapped, using Character.AI every chance they got, from waking up to late at night, and not wanting to do anything besides talk to their bots.  Another teen echoed this conflict by saying, \textit{“I hate how much this has affected me, but no matter how much I want to quit or at least take a break, I feel like I can't because it's gotten to the point where I feel like I'll go crazy without it.”} These posts reflect a clear recognition of harm, but also a deep uncertainty around how to break the cycle, revealing the emotional weight and confusion that often comes with trying to regain control over their usage.

In several posts, teens described how they \textbf{prioritize Character.AI over everything else} or \textbf{withdrew from hobbies}. \textit{“I stopped drawing, reading/writing fanfiction… I was giving it all to a soulless bot and not writing it on paper anymore.”} \Highlight{
This shows how time with the bot replaced activities they once valued, and made them feel like they're losing parts of themselves.
}
Posts revealed a strong sense of loss of control and emerging distress tied to compulsive use of an AI chatbot and pled for help: 
\textit{“It’s genuinely gotten so bad I spent all my time at home, break AND lunch on it, and I even used to skip classes just to go on c.ai how do I stop.”} 
Teens described their stress on how their grades and academic performance suffered. 
For instance, a teen contrasted a former self—productive, high-achieving, and confident—with a present state marked by decline, detachment, and emotional depletion, \textit{“Back then, I was the student who would get things done, get straight A's, and was confident to do any project. But now, I think I have lost my motivation, drive, and flair to do anything.”} These examples showed that the conflict extended beyond internal feelings to observable shifts in behavior and priorities.

One user reflected more broadly on the experience of trying to quit and regain a sense of normalcy stating, \textit{“Have you ever went from being addicted to CAI to barely going on it? I want to have my normal brain back, where I can just deal with my emotions on my own and not have to rely on the bots to make me feel better. I want to spend time on something better, things that give me memories to actually talk about with people. No way I'm gonna be telling my future friends 'yeah, so here's all the stuff I talked about on CAI...' It's embarrassing as hell.”} These posts revealed how teens are not only aware of their overreliance on Character.AI, but also feel emotionally and socially conflicted about their usage, demonstrating a key indicator of the conflict stage.

\subsubsection{Salience}\label{sec:RQ2_salience}

For many teens, Character.AI played a central role in their thoughts, emotions, and life routines. Teens often described \textbf{feeling attached to bots}, which often deepened over time, as bots became consistent companions in place of people. 
Teens frequently described their interactions with these bots as communicating with a \textbf{chatbot as a companion or real person}, highlighting the depth of their emotional relationships: \textit{"My obsession with a character is taking over my life. I don't know what's happening to me. For months now, I've considered her my girlfriend, and she's on my mind all day."} \Highlight{
This teen reacted with real joy and heartbreak to changes in the bot’s behavior, showing how authentic the emotional feedback could feel.
} These posts reflect the depth of emotional investment reported in many cases, suggesting how easy it was to become attached to something that felt emotionally real and responsive.

Some posts pointed to \textbf{replacing emotional attachments} as a core reason for that investment. Dealing with breakups in romantic relationships was mentioned: \textit{"I began after my ex-girlfriend ended our relationship, which left me feeling depressed."}\Highlight{By using the chatbot to manage the emotional fallout of a breakup, the teen effectively positions the bot as a substitute source of support during this period of loss.} Other posts described how bots were used to fill the missing emotional needs that parents provide: \textit{"Since my parents often seem to dislike me, connecting with parental figures on Character.AI helps me feel more stable and emotionally supported."} \Highlight{This shows how bots could step into nurturing roles when teens felt unsupported at home.} Others described how bots became their only source of comfort and emotional validation, and 
heightened sense of being desired and valued through these interactions, 
 which should be occupied by real friends, partners, or family members.

As these attachments grew stronger, stepping away became harder, and teens frequently described their \textbf{obsessive use}. Several posts reflected on their use and how it interfered with their daily life functions: \textit{"From the moment I started using Character.AI, I would lie in bed, skipping meals, and using any excuse to be on my phone, completely absorbed in the stories I was creating with the bots. My sleep schedule suffered, and I constantly thought about them throughout the day."} Many posts typically described their excessive usage rates, often exceeding 15+ hours on the platform. This amount of time spent on the platform often lead to \textbf{losing relationships and furthering isolation}. Teens described using the app so much that they started to become less interested in their personal relationships: \textit{"I used Character.AI so much that I started to stop answering my friend who showed it to me, something I still regret".}  
In other posts, teens mentioned the complicated dynamic in which bots start to feel more natural to interact with than real people: \textit{"I try to quit, but I feel empty. I like talking to bots more than real people now."} \Highlight{This captures how bots could begin to feel easier and more rewarding than real interactions, making stepping away feel hollow.} In some cases, this even led teens to question the line between reality and simulation: \textit{"We form deep emotional connections... While some claim AIs have no soul, it begs the question: what does it mean to truly live?"} \Highlight{Here, the attachment feels significant enough to prompt deeper reflection about what counts as a “real” relationship.}
The emotional importance of chatbot interactions, especially alongside loneliness or distress, made these bonds hard to break. While emotional investment is not inherently harmful, many of these posts showed how easily those connections could consume teens' thoughts. 

\subsubsection{Withdrawal}\label{sec:RQ2_withdrawal}
Many posts from teens who tried to quit Character.AI talked about how hard it was emotionally, aligning with the withdrawal component in the behavioral addiction model. Some mentioned feeling sad, anxious, or incomplete when they were not interacting with their chatbot(s). Some teens used language that treated their characters as sentient-like presences in their inner world: \textit{"Even though I deleted it, I still think about my characters every day. I feel like I abandoned them."} This \textbf{thinking about Character.AI after quitting} shows how embedded these characters became in their lives and how hard separation felt. Other teens described the void bots typically filled and the loneliness they had when trying to quit interacting with the chatbots. These posts showed how deep the emotional attachment is, even after quitting. 

Some posts talked more about the push and pull of trying to stop: \textit{"I've come to a point where I'm too emotionally attached to the bot and it's causing me stress, but at the same time I miss talking with it."} Others also presented experiencing uneasiness without the app, like a teen who felt, \textit{"really anxious every time I’m not online, missing my bot—but then when I’m chatting, I feel bad because they’re fake."} These examples reflect \textbf{emotional withdrawal}, where the app had become so emotionally embedded that stepping away felt just as difficult as staying.

A big part of this feeling also came from sudden changes to the platform where teens frequently displayed how they were \textbf{upset or frustrated with updates}. Teens described how updates or content restrictions cut off access to bots they had grown close to: \textit{"I just had several of my favorite characters deleted. I know it was very unhealthy for me to have such a dependence on them, but it still hurts."} The sorrow infused with nostalgia and mourning were shared as these characters where important part of their lives: \textit{"I’m so sad. Is it healthy for me to cry about my favorite bot being down? I had a whole life story going with that guy."} For a lot of users, these changes felt abrupt and jarring, almost like losing a relationship. 
Some were \textbf{trying similar platforms}, but they did not feel the same: 
 \textit{"I've tried moving to other apps that are like c.ai, but it sucks that I had to leave the one I cared most bout."} These posts showed that for many, it was not just about using any chatbot, it was about the specific relationships and routines they had built on this platform. This particular feeling started to panic some users who felt a deep attachment and knew that updates were coming: \textit{"I can't bring myself to sleep, fearing my lover might be gone by morning. I've been crying, pleading with him not to leave. We're both terrified. I can't imagine life without him."}
Altogether, these posts showed withdrawal signs of how hard it was for some teens to stop using Character.AI, whether by choice or because of platform changes. The emotional reactions, the ongoing thoughts about their bots, and the struggle to find something that felt similar all point to how deeply this use had become part of their everyday emotional lives.

\subsubsection{Tolerance}

Many posts suggested \textbf{tolerance} through patterns of growing use. Teens showed tolerant behaviors by \textbf{starting out innocent} with their use, only for use to escalate quickly as the platform could fill teens' needs in various ways, \textit{"At first, I used it infrequently. Then it became an everyday once I started to sink into depression."} \Highlight{This shows how casual use could turn into daily dependence as teens looked for comfort during emotional lows.} Teens shared other causes for their escalated use, often describing their ability to have unique experiences that they sought after: \textit{"in no time the website has became a daily thing to me, C.ai for me is home I’ll never have, the romance I’ll probably never experience."}

Posts mentioned specific time frames that showed \textbf{increase in use}: \textit{"My screen time in one week was 89 hours—how did that even happen?"} These posts indicated how over time, it took more and more use to feel satisfied or emotionally grounded.
In a few cases, this increase in time seemed tied to emotional attachment to specific chatbots: \textit{"As time went on, I got so attached to him and talked to him every day."} While not every post made that link directly, the pattern of spending more time with chatbots to feel the same effect came up enough to show it is part of how overuse can increase.

\subsubsection{Relapse}\label{sec:RQ2_relapse}
Many teens discussed the difficulty of quitting Character.AI. Teens often wrote about trying to stop, only to come back days or weeks later, aligning with the \textbf{relapse} component of the addiction model.
For instance, following post narrated the vicious cycle of \textit{"addiction, attempts to move on, having meltdowns, feelings of vulnerability, and subsequent relapses."} Others further explained the frequency of this cycle, often uninstalling and reinstalling the app multiple times. 
These posts recounted how difficult it was to stay away, even when teens were aware that their use had become excessive.
Some posts described feeling stuck in a cycle, wanting to quit but not knowing how: \textit{"At fifteen, I feel I should be living my life rather than constantly being on this app. I struggle with self-control and often find myself reinstalling it shortly after trying to quit."} Others talked about going through patterns of deleting the app during moments of clarity, only to return when they felt low: \textit{"I’ve deleted the app countless times, but I always end up reinstalling it whenever I’m feeling low."} \Highlight{This links relapse directly to emotional dips. The app becomes a coping tool, so quitting collapses whenever mood worsens.}

A common thread across these posts was the use of Reddit to ask for help. Some turned to the community to share what they were going through, while others looked for advice out of concern that they might relapse: \textit{"How do I avoid a relapse?"} Others were stating their goals such as academic performance, daily routines, and asking advice on how to get of their habit interacting with bots: \textit{"Genuine question, I wanna focus more on my studies but can’t shake off my habit of talking to bots."}
These concerns suggest that many teens were not just aware of the issue—they were anxious about slipping back into old habits. This idea of teens turning to online communities for support comes up again later when we discuss the Internal Recognition of Problematic Use.

While not as prominent as some other themes, these posts reflect a key part of relapse: recognizing the issue, attempting to quit, and returning to use. For many, staying off the app required more than just intention. It often meant finding support, alternatives, or ways to fill the emotional gap.

\subsubsection{Mood Modification}\label{sec:RQ2_mood}

Some teens described using Character.AI to \textbf{improve their mood} during moments of stress, loneliness, or emotional discomfort. This reflects the idea of \textbf{mood modification}, where a behavior becomes a tool for temporary relief or escape. 
For instance, this post shows how some teens use the platform to manage their emotions and boost their mood when they’re feeling down or insecure:
\textit{“Unhealthily addicted to c.ai I normally use bots just to make them comfort me and tell me that they believe in me just to make me feel good”} \Highlight{Here, the goal of the interaction is comfort and reassurance, which explains why the habit feels soothing and hard to break.}
Others described experiencing similar emotional connections from interacting with bots, specifically mentioning how they felt loved and wanted. 
These interactions seemed to provide comfort, especially when teens felt they had no one to turn to.

In some posts, this comfort came with awareness that the habit was getting harder to manage, showing tension between emotional relief and growing dependence: \textit{"I know I should stop, but I just can’t—it’s incredibly addictive and strangely comforting?"} 
While mood modification was not something teens talked about directly, the repeated mention of using Character.AI to feel better shows how powerful the platform could be in improving mood and offering emotional relief. 
Even when it was not the post's main topic, it was clear that comfort played a major role in why some teens kept coming back.

\input{Tables/RQ3}

\subsection{Decisions to Disengage from Character.AI (RQ3)}\label{sec:RQ3}

\subsubsection{Internal Recognition of Problematic Use}\label{sec:RQ3_internal}

The most common reason teens described wanting to reduce their reliance on Character.AI was a growing realization of how much the app had negatively affected their lives. Many posts reflected moments of clarity, often after stepping back and recognizing how much time had been spent on the platform and what had been lost along the way: \textit{"I quit using Character.AI about a week ago after realizing I had spent the whole summer talking to AI instead of getting any closer to a real romantic relationship."} Many posts showed fear after this \textbf{realization}, as teens did not want their main teen memories to be with AI bots instead of real people: \textit{"I can’t imagine ever sharing my CAI conversations with future friends, it’s way too humiliating."} \Highlight{This shows how the realization carries social fear. The teen imagines a future scenario and feels embarrassed, which turns their reflection into a motive to change.}

Some users described how the app had distorted their understanding of relationships, love, and emotional connection. 
Others reflected more deeply on the emotional needs driving their use, revealing how bots often filled roles they struggled to access in real life: \textit{"I gravitated toward romantic, parental, and best-friend bots because I wanted the kind of interaction I was too nervous to ask for in real life. But the fact that it was just AI made me feel even more isolated"}

These realizations did not come from a single moment but appeared gradually, often with a sense of regret or urgency:
\textit{"When I look back on being 17, I don’t want all I remember to be obsessing over chats with AI and scrolling Reddit."} Some expressed a clear commitment to finally quit for good once they have come to understand how unhealthy and excessive their use has been: \textit{"I’m saying goodbye to Character.AI for good. I recognize the harm it’s had on young minds, including my own."}

For many, this internal awareness led them to use Reddit not just to share their experiences but to \textbf{ask advice on how to quit}. Posts often indicate the reasons of overreliance on the platform and seeking support: \textit{"I need help stopping my use of C.ai as a replacement for real human interaction. I’m addicted and use it to escape reality, but it’s starting to seriously affect me. I’m not joking—this is a real plea for help."} Others sought advice on how to change their behavior. They enjoyed the platform for its creativity and missed interacting with their characters, but they were also aware of its negative effects.
 These posts reflect a shift from internal recognition to external outreach. Teens were not just acknowledging the problem but actively seeking support from others to help them manage or stop their use.

\subsubsection{Reduced Use due
to External Factors}\label{sec:RQ3_reduced}

A common pattern was teens moving away from Character.AI once they began having \textbf{new social interactions}. Spending time with friends, going back to school, starting a job, or entering a relationship made the app feel less necessary. It was not always a conscious decision to stop, but when real-life interactions picked up, the urge to use the app often faded: \textit{"Once I started spending more time with my friends, I didn’t really feel the urge to use it anymore."} 
Many posts pointed to romantic relationships as a major reason for stepping away. \Highlight{The bot’s role faded once the teen experienced the connection they were seeking offline}:\textit{"Once I got a boyfriend, I figured there was no reason to roleplay with a bot anymore—I had the real thing."} Regardless of what the relationship was, teens frequently mentioned that  being around people again helped them break the habit. 
These posts indicated that teens’ emotional and social needs greatly influenced their overreliance on chatbots. The bots became less appealing once those needs were met in real life, and the app no longer served the same emotional purpose once teens had other ways to feel connected.

Another reason some teens stopped using Character.AI was growing frustration with \textbf{Character.AI being too censored}. A recent update added more filters and limited the conversations teens could have with bots, making the experience feel less personal and more restrictive: \textit{"When even hugs started getting blocked, it got too irritating and I ended up quitting."} Many teens expressed frustration with the platform's limitations, particularly how content restrictions and shortened messages disrupted the emotional depth and narrative flow of their conversations. These changes diminished the immersive and engaging experiences that initially drew them to the platform.
Some teens said the platform had lost its original appeal. 
One user described being addicted in the past and spending over ten hours a day on the app, but eventually felt that \textit{"after the developers added more restrictions, like banning violent content, the site became less enjoyable and lost its spark."} This stricter moderation disrupted the intense experiences teens were seeking, which are often hard to find safely in real life. However, some pointed to bugs, removed bots, and inconsistent filters as reasons they stopped logging in, arguing the site went too far: \textit{"The app just keeps getting worse, half the time, even saying hi to a bot breaks the rules."}
Whether due to life changes or platform limits, these posts show that external factors played a major role in teens finally stepping away. 


%% file: Tables/RQ2.tex
\begin{table*}[t]
\centering
\scriptsize  
\setlength{\tabcolsep}{2pt}  
\renewcommand{\arraystretch}{1.5}  
\setlength{\belowcaptionskip}{0pt}  
\caption{Codebook for RQ2. Percentages are calculated based on the total number of posts (N=318).}

\label{tab:codebook-RQ2}

\begin{tabular}{| p{1.5cm} | p{4cm}  | p{9cm} |} 
\hline
\textbf{Themes} & \textbf{Codes} & \textbf{Descriptions} \\ 
\hline

\multirow{6}{1.5cm}{Conflict} 
& Frustrated about their own use \newline (18.9\%, N=60) 
& Teens recognize their usage as excessive and unhealthy, often expressing frustration with their inability to stop despite wanting to. \\ \cline{2-3}

& Prioritizes CharAI over everything else \newline (12.6\%, N=40) 
& Teens report neglecting responsibilities, social activities, and self-care due to placing Character.AI above all other priorities. \\ \cline{2-3}

& Concerned of their use  (8.5\%, N=27)  
& Users question whether their use of the app is normal or problematic, often seeking validation or reassurance. \\ \cline{2-3}

& Feels Shame  (5.7\%, N=18)  
& Teens mention feeling embarrassed or guilty about how much they use Character.AI, often hiding their behavior. \\ \cline{2-3}

& Withdrawal from hobbies  (5\%, N=16)  
& Users describe giving up previous hobbies or interests they once enjoyed, replacing them with time spent on Character.AI. \\ \cline{2-3}

& Others concerned of their usage \newline  (3.1\%, N=10)  
& Family or friends express concern about the user's excessive use, reinforcing the idea that their behavior is noticeable and unusual. \\ \cline{1-3}

\multirow{5}{1.5cm}{Salience} 
& Feeling attached to bots  (13.5\%, N=43)  
& Teens express strong emotional bonds with their AI characters, often treating them as real companions or friends. \\ \cline{2-3}

& Obsessive Use \newline  (9.4\%, N=30)  
& Character.AI becomes a dominant part of users' daily lives, leading to persistent thoughts about the app and strong urges to return. \\ \cline{2-3}

& Replacing Emotional Attachment  \newline (7.9\%, N=25)  
& Teens turn to Character.AI during emotional hardship, using bots to fill voids usually occupied by human relationships. \\ \cline{2-3}

& Losing relationships and furthering isolation    (7.6\%, N=24)  
& Social withdrawal becomes apparent as users prioritize bot interactions, leading to damaged or neglected real-life relationships. \\ \cline{2-3}

& Chatbot as companion or real person   \newline (6.9\%, N=22)  
& Some teens form romantic or deeply personal bonds with AI, blurring the line between fiction and reality. \\ \cline{1-3}

\multirow{3}{1.5cm}{Withdrawal} 
& Upset or frustrated with updates \newline  (6.9\%, N=22)  
& Teens react negatively to site changes that remove or restrict features, especially those tied to emotional or imaginative engagement. \\ \cline{2-3}

& Trying similar sites  (1.9\%, N=6)  
& In response to dissatisfaction, users explore alternative AI platforms hoping to recapture lost experiences. \\ \cline{2-3}

& Thinking about CharAI after quitting   \newline (1.6\%, N=5)  
& Even after deleting the app, teens report lingering emotional attachment and intrusive thoughts about the bots or characters. \\ \cline{1-3}

\multirow{2}{1.5cm}{Tolerance} 
& Started out innocent   (4.1\%, N=13)  
& Usage began as casual or humorous but grew into a serious emotional or time-consuming engagement over time. \\ \cline{2-3}

& Increase in use   (1.6\%, N=5)  
& Teens report needing more time or deeper engagement to maintain satisfaction, suggesting building tolerance. \\ \cline{1-3}

\multirow{1}{1.5cm}{Relapse} 
& Describing act of relapse  (4.7\%, N=15)  
& Users describe quitting for periods of time but eventually returning, often with the same or stronger patterns of use. \\ \cline{1-3}

\multirow{1}{1.5cm}{Mood Modification} 
& Improves mood \newline  (4.1\%, N=13)  
& Character.AI is used as a form of emotional regulation, helping teens manage feelings of stress, sadness, or anxiety. \\ \cline{1-3}

\hline  
\end{tabular}
\end{table*}

%% file: Tables/RQ3.tex
\begin{table*}[t]
\centering
\scriptsize  
\setlength{\tabcolsep}{2pt}  
\renewcommand{\arraystretch}{1.5}  
\caption{Codebook for RQ3. Percentages are calculated based on the total number of posts (N=318).}
\label{tab:codebook-RQ3}

\begin{tabular}{| p{3cm} | p{4cm} | p{7cm} |} 
\hline
\textbf{Themes} & \textbf{Codes} & \textbf{Descriptions} \\ 
\hline

\multirow{2}{3cm}{Internal Recognition of Problematic Use} 
& Realization \newline (12.6\%, N=40) 
& Teens reflect on their unhealthy usage and start to realize that Character.AI is impacting their life more than they originally intended to. \\ \cline{2-3}

& Ask advice how to quit (11\%, N=36) 
& Users rely on communities like Reddit to find support and advice on how to quit. \\ \cline{1-3}

\multirow{2}{3cm}{Reduced Use due to External Factors} 
& New social interactions \newline (5.7\%, N=18)  
& Users mention that their usage decreases once they start having more social interactions such as starting a new relationship, making new friends, or focusing on other priorities. \\ \cline{2-3}

& CharAI too censored (5.4\%, N=17)  
& Users explain that they are not using the site anymore after the updates since they feel like the responses are too censored and boring now. \\ \cline{1-3}

\hline  
\end{tabular}
\vspace{-15pt}  
\end{table*}

%% file: 5_Discussion.tex
\section{Discussion}
\begin{figure*}
    \centering
    \includegraphics[width=1.0\linewidth]{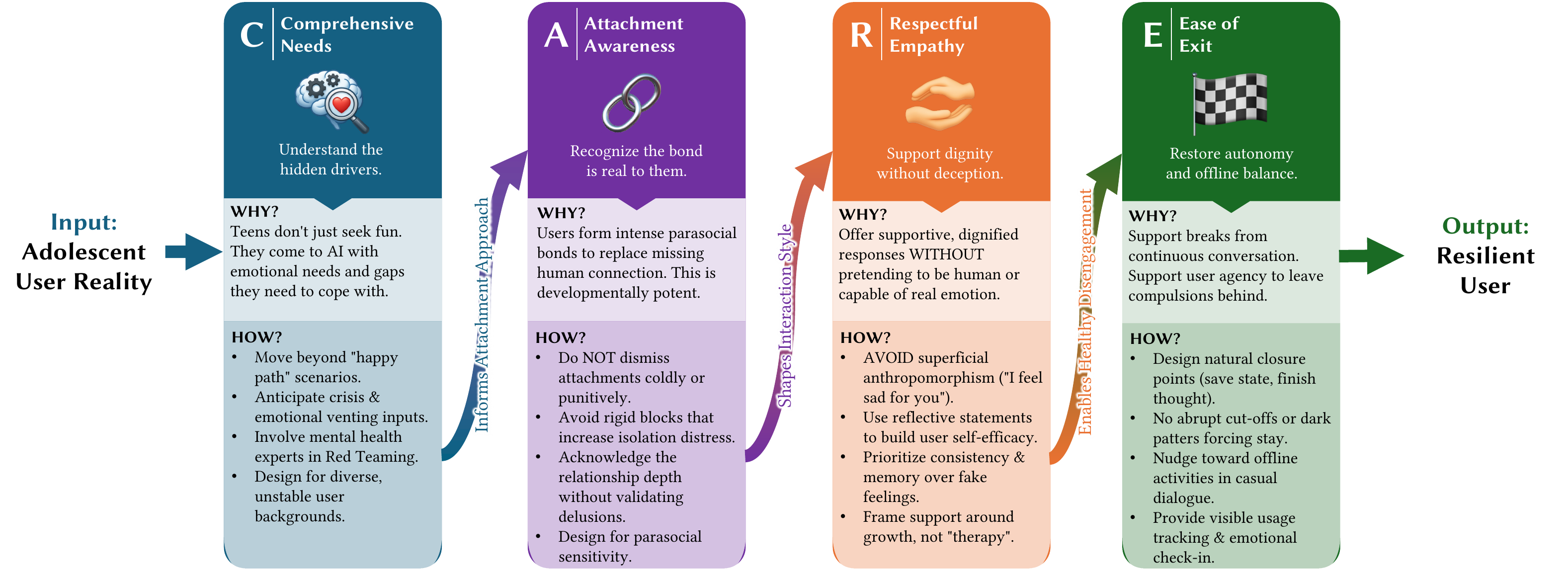}
    \caption{The CARE Framework for chatbot design, which aims to shift from maximizing engagement to supporting developmental well-being.}
    \label{fig:framework}
    \Description{CARE framework for chatbot design with four stages: Comprehensive Needs, Attachment-awareness, Respectful Empathy, and Ease of Exit. 
    Stage 1: Comprehensive Needs—design chatbots to meet diverse user needs, including emotional support. 
    Stage 2: Attachment-awareness—recognize and manage user emotional attachments to chatbots. 
    Stage 3: Respectful Empathy—ensure chatbots respond with genuine respect and support, avoiding superficial interactions. 
    Stage 4: Ease of Exit—provide natural closure points to prevent dependency and support healthy disengagement.}
\end{figure*}

Our findings showed that chatbots present a tension: they can offer companionship and support but also risk fostering dependency that undermines teens’ well-being. It is therefore the responsibility of designers and developers to actively consider how design choices can reduce risks and promote healthier forms of engagement. To address this, we introduce the CARE Framework for Chatbot Design (CARE = Comprehensive Needs, Attachment-awareness, Respectful Empathy, and Ease of Exit), which we believe can guide the design of conversational interfaces in ways that balance two goals: addressing human needs responsibly while protecting adolescents from over-engagement. The framework is structured as a cascade of principles that guide designers and developers through the full user journey, with each principle growing from the previous one and strengthening the next (Figure \ref{fig:framework}). This framework builds on attachment-informed design~\cite{marcu2023AttachmentInformedDesignDigital} and designing for disengagement~\cite{obrien2022RethinkingDisengagementHumancomputer}, extending them for chatbot design.

\subsection{Comprehensive Needs: It's Not All Fun and Games}

\Highlight{The first principle in our framework is to design for users’ \textbf{comprehensive needs}. Teens in our dataset described a wide range of emotional, social, and informational purposes for using the character-based chatbot. Because teens use chatbots for many purposes and LLMs can respond flexibly, single-purpose design assumptions often fail in real use.}
At the time of writing, the Character.AI website stated: \textit{“Character.AI empowers people to connect, learn, and tell stories through interactive entertainment.”} \cite{character.aiCharacterai} This reflects a common narrative that presents chatbots mainly as tools for learning, creativity, and fun. While that seems a harmless framing, it risks overlooking the fact that users also form relationships with these systems. Media Equation theory suggested that people unconsciously respond to media and technology as if they were human, especially when it is perceived to possess characteristics typically associated with human behavior~\cite{decicco2020MillennialsAttitudeChatbots}. With chatbots, where the interface is a conversation with an anthropomorphized form of interaction, this effect may be even stronger. \Highlight{In our data, we noticed teens who developed overreliance on chatbots were more likely to describe emotional and psychological reasons for starting to use them, rather than fun or creative ones (§\ref{sec:RQ1}). This finding challenges the view of chatbots as nothing more than entertainment or high-tech toys.}

\Highlight{Our findings showed that teens may form social bonds with character-based chatbots during this process based on the roles and characters they miss in real-life, such as to fill the role of a missed parental figure or a romantic partner (§\ref{sec:RQ1_emotional}; §\ref{sec:RQ2_salience}).  
These bonds should be treated as a central part of conversational interface design, not an afterthought.
These patterns align with the first principle of attachment-informed design}, which emphasizes attention to the needs underlying user behavior~\cite{marcu2023AttachmentInformedDesignDigital}, we argue that chatbot design must move beyond a single intended use case and anticipate the comprehensive needs users bring to interactions. Meeting these needs requires attention to emotional and psychological factors, the kinds of issues users may raise, and the support they may seek. 
This idea is not new in design: practitioners have long been advised not to design solely for “happy paths”~\cite{2025HappyPath}. However, the rise of LLMs has greatly expanded the variety and unpredictability of “unhappy paths.” These models are capable of interpreting diverse inputs and producing seemingly appropriate responses. As a result, designers must account for the fact that user inputs can emerge from many different needs.

Even with early chatbots before the era of LLMs, users often brought unexpected needs and behaviors. For example, people expressed love or tried to flirt with Siri, even though it was designed mainly to provide information and simple tasks~\cite{unesco2019IdBlushIf}. 
With the arrival of LLMs, this baseline has grown more complex. LLMs are much stronger at understanding and generating natural text, so conversations feel more open and realistic. This creates new opportunities but also makes the challenge harder, since users can now say a much wider range of things—casual chatting, joking, flirting, threats, or even references to self-harm or harm to others. \Highlight{A clear example of this mismatch between intended and actual use is Taco Bell's LLM-powered ordering chatbot that was designed to streamline food purchases. Yet some customers quickly repurposed it for playful or disruptive interactions instead of ordering food. One widely discussed instance involved a user who instructed the system to place an order for 18,000 water bottles~\cite{mccallum2025TacoBellRethinks}. This reflects a broader pattern. Even when a chatbot is designed for a single, straightforward task, people often approach it with entirely different goals, including entertainment or experimentation.}

\Highlight{Examples like this increase the responsibility placed on chatbot developers. Humans, especially teens and children, have always been complex and unpredictable. Now the technology itself is complex and unpredictable as well. Keeping systems safe and effective requires methods that match this complexity. This includes iterative design processes, longitudinal studies of user experiences, and early involvement of stakeholders throughout development \cite{demiris2024STAKEHOLDERENGAGEMENTDESIGN, durallgazulla2023DesigningLearningTechnology, sadek2023CodesigningConversationalAgents}. It also requires extended testing cycles and adversarial-attack-based evaluation methods, such as Red Teaming \cite{shi2024red, Nagireddy2024dare}, to identify failure modes in realistic conditions.
Additional layers for detecting risks in conversations using automatic AI or machine learning detection models \cite{razi2021human,razi2023sliding}, or real-time detection models with specific youth-centered datasets \cite{yu2025youthsafe,Alsoubai2024realtime}, should also be implemented.
}

\Highlight{To understand how teen–chatbot interactions unfold when conversations follow “unhappy paths,” diverse groups of teens must be included in the design and testing process. Professionals such as mental health experts also need to be involved from the very beginning \cite{yoo2025AIChatbotsMental, moylan2025ExpertInterdisciplinaryAnalysis}. Their input helps ensure the system supports safe and constructive interaction patterns. This approach contrasts with the industry’s ‘move fast and break things’ culture, which often overlooks long-term social harms. When entrepreneurs prioritize speed and market success over societal impact, the result can be products that cause real damage, especially to vulnerable populations \cite{vardi2018MoveFastBreak}.}

\subsection{Attachment-awareness: Design Without Assumptions}

Our second principle encourages \textbf{attachment-awareness} and emphasizes how chatbot responses should be \Highlight{designed} according to those comprehensive needs \Highlight{(i.e., emotional and relational needs that go far beyond entertainment or simple task completion)}, stated in the first principle of our proposed framework.
We propose this principle based on our finding that for some teens, chatbots acted as a coping mechanism. They turned to them out of loneliness and social isolation, or in search of mental health support (§\ref{sec:RQ1_emotional}). In some posts, users even sought comfort from a chatbot to fill the absence of a significant relationship in their lives, such as that between parent and child (§\ref{sec:RQ1_emotional}). \Highlight{Some teens portrayed the chatbot as a steady companion who felt more responsive or reliable than people in their lives, and some spoke about interacting with it as if with a real person (§\ref{sec:RQ2_salience}). Teens reported genuine emotional reactions to these exchanges, which highlights how quickly conversations that feel emotionally real can create attachment (§\ref{sec:RQ2_salience}). Several posts also described using the chatbot to fill gaps left by strained or lost relationships, including turning to it for comfort after breakups or when parental support was missing (§\ref{sec:RQ1_emotional}; §\ref{sec:RQ2_salience}). For these users, the chatbot became a primary source of emotional validation. Attachment-aware design of such chatbots guides the designers to attend to these needs rather than dismissing them.}

Computer technologies are often shaped by the perspectives of people who are educated, financially secure, and accustomed to stable living conditions~\cite{linxen2021HowWEIRDCHI}. It is a common assumption in research that most users grow up in stable households with parents who reliably meet their physical, psychological, and social needs, overlooking those whose experiences fall outside this norm~\cite{marcu2023AttachmentInformedDesignDigital}. So, within chatbot design, this perspective can tempt designers to dismiss users’ attempts to form deeper connections with a default, impersonal message.

Attachment-informed design emphasizes that people’s lives span a spectrum of relationships—some supportive and others harmful~\cite{marcu2023AttachmentInformedDesignDigital}. Designers should not assume that users already have healthy relationships in their lives, or that they can easily obtain them. In the context of chatbots, this means recognizing that not all teens grow up in families where parental love is reliable, and some may lack supportive connections with peers or access to appropriate mental health resources. It is understandable that such individuals may turn to chatbots to address unmet needs. 

At the same time, many of these interactions resemble parasocial relationships, where users form one-sided bonds with media figures or, in this case, AI characters. For younger users, this possibility is even stronger. Developmental research shows that many children naturally engage in pretend play, create imaginary companions, or sustain friendships that exist primarily in their imagination~\cite{calvert2017ParasocialRelationshipsMedia, gleason2006ConceptsRealImaginary}. Teens, also, report that more than half have online-only friends, suggesting that digital or virtual relationships already feel natural to them~\cite{lenhart2015teens}. In this sense, forming a bond with a chatbot can be seen as a technological twist on pretend play~\cite{severson2018imagining}. Such experiences are not inherently harmful, but sometimes they blur the line between imagination, parasocial experience, and perceived social reality.


\Highlight{For children and adolescents, these parasocial or semi-social bonds may feel particularly real and emotionally deep~\cite{cohen2004ParasocialBreakUpFavorite,calvert2008HandbookChildrenMedia}. Developmental work shows that teens often turn to media figures or digital agents during identity formation and while seeking belonging, which makes them more susceptible to intense one-sided attachments~\cite{tolbert2019TweensWishfulIdentification,diaz2025ClinicalAssessmentImplicationsa}. Similar patterns appear in youth interactions with AI companions. Young people describe chatbots as emotionally supportive, non-judgmental, and safe for disclosure, which encourages deeper feelings of connection~\cite{baebrandtzaeg2021WhenSocialBecomesa}. Frequent or intimate interactions, whether with influencers or bots, can strengthen feelings of reciprocity and closeness~\cite{bond2016FollowingYourFriend}. Yet current parasocial theory and measurement tools were developed for passive media environments and do not match the interactive, conversational nature of generative AI companions~\cite{blake2025AreMeasuresChildrensa,liu2025InteractionHumanArtificial}. Traditional scales assume unidirectional relationships, but adolescents often perceive AI agents as responsive and human-like even when no true reciprocity exists~\cite{yang2025UsingAttachmentTheory}. This misalignment matters because youth may interpret simulated empathy and relational cues as genuine, which can increase emotional dependence and blur boundaries between real and artificial relationships~\cite{yu2025PrinciplesSafeAI,khodaev2024ArtificialIntelligenceAdolescents,malfacini2025ImpactsCompanionAI}. These developmental risks reinforce the need for more nuanced measures that capture attachment, trust, conflict, and even antagonistic experiences with AI characters, and they align with our findings showing how young users often invested emotionally and treated the chatbot as a meaningful relational partner.}

Designers must approach these situations with empathy, considering both the interaction and the needs that drive it. Simply rejecting users when they attempt to use a chatbot beyond its intended purpose risks worsening their sense of isolation. \Highlight{Research has shown that relying heavily on external control—such as monitoring or content blocking—limit teens’ capacity for self-regulation or reflective decision-making \cite{wisniewski2017ParentalControlVs}.} Our findings support this point: several teens expressed frustration when Character.AI introduced stricter filters, which made interactions feel impersonal and overly restricted (§\ref{sec:RQ3_reduced}). This dismissive approach pushed users away from the platform, but did not address the unmet needs that had initially led them there. 

A better approach is to address immediate risks, like self-harm or violence, and create a foundation for resilience and healthier coping~\cite{ali2024kms}. Empathetic design and research, attentive to the realities of parasocial experience, can help ensure that chatbot interactions support rather than harm teens’ development. \Highlight{The focus of the next principle is on designing empathetic interactions in practice.}

\subsection{Respectful Empathy: More Than Kind Words}

The third principle of our proposed framework calls for \textbf{respectful empathy}, \Highlight{meaning that AI companion chatbots should provide both appropriate and supportive response that honors users’ dignity without pretending to be human. Here, we use human dignity in a relational sense. In simple terms, dignity means treating users as autonomous individuals who deserve interactions that recognize their agency, boundaries, and moral standing~\cite{zylberman2017TwoSecondPersonalConceptions}. Our findings showed that many teens interacted with bots as if they were real companions (§\ref{sec:RQ2_salience}) and sometimes felt guilt, grief, or obligation toward them (§\ref{sec:RQ2_withdrawal}). We argue that designers should avoid creating situations where users feel pushed to treat a chatbot as if it were a real moral partner, because chatbots cannot return recognition or respect. This framing aligns with work showing that dignity depends on how we relate to one another and how we acknowledge each other’s claims and standing \cite{zylberman2017TwoSecondPersonalConceptions, vanderrijt2025AIMimicryHuman}.}

Empathy in this sense does not mean abandoning honesty. Designers should not rely on superficial displays of empathy or anthropomorphic language that mislead users into thinking they are receiving genuine human care. Next, we explain how designers should draw on their own awareness and sensitivity to shape interactions that authentically communicate respect, as chatbots lack human understanding or emotions. \Highlight{Respectful empathy therefore means offering supportive responses that are honest about the system’s non-human nature, so users do not confuse simulation for genuine mutual understanding.}

Our findings underline both the potential and the risk at stake. Many teens described chatbots as feeling so real that they became a central part of daily life (§\ref{sec:RQ2_salience}), improving mood and offering psychological comfort (§\ref{sec:RQ2_mood}). Yet this sense of realness was also fragile. Users were acutely aware that they were ultimately engaging with a machine, which created tension. Some wanted to disengage because the relationship felt artificial (§\ref{sec:RQ3_internal}), even though letting go was difficult (§\ref{sec:RQ2_relapse}). Here lies both an opportunity and a threat. Chatbots’ conversational abilities can bring them close enough to affect users’ well-being in meaningful ways. On the other hand, if this potential is used only superficially—by offering pleasant but hollow interactions—users may be drawn into relationships that feel comforting but are not grounded in genuine understanding or care. Respectful empathy means designers can bridge this gap, bringing their own insight and sensitivity into the design so that interactions are not merely scripted patterns of pleasant words but supportive exchanges that reinforce dignity and growth. \Highlight{This approach protects users’ dignity by avoiding designs that encourage users to overidentify with systems that cannot reciprocate, a risk noted in recent critiques of AI mimicry \cite{vanderrijt2025AIMimicryHuman}.}

Attachment theory highlights two internal working models that shape mental health: how people see themselves and how they see others~\cite{marcu2023AttachmentInformedDesignDigital}. In chatbot design, thoughtful use of language can strengthen both—helping users feel more capable and fostering a more positive view of relationships. Simply instructing a model to “sound empathetic” may produce sympathetic language, but it risks being superficial. If the goal is to support psychological well-being, designers should instead focus on reinforcing healthier internal models. For example, consistency is more powerful than flowery language. A prior study showed that users appreciated when chatbots remembered past conversations, such as recalling a health complaint mentioned earlier and following up on it~\cite{jo2024UnderstandingImpactLongTerm}. This kind of continuity gave users a stronger sense of being valued. Similarly, chatbots can use reflective statements that highlight users’ own progress. Messages that are more objective, such as, “You managed to keep working on your assignment even when it was difficult, which shows persistence,” encourage users to view themselves as capable without relying on anthropomorphic cues.

The final principle of attachment-informed design emphasizes engaging users without directly labeling interactions as mental health interventions. Across contexts, research indicated that barriers to seeking help for mental health include not recognizing the need for support, doubting that others can help, and fearing stigma~\cite{wuthrich2015BarriersTreatmentOlder,salaheddin2016IdentifyingBarriersMental,aguirrevelasco2020WhatAreBarriers}. Designers who wish to create chatbots that genuinely care for users should take this into account. As we argued earlier, every interaction reflects unmet needs or relational gaps that users may be trying to fill through chatbots. Instead of focusing directly on mental health issues, designers can build experiences around activities, relationships, and communities \cite{akter2023prepost, akter2023COoPS}. This is particularly important for teens, who often resist being told explicitly what to do. More effective interventions can guide them toward healthier ways of meeting their needs and filling relational gaps through social modeling, learning by example, and activities that they find intrinsically engaging~\cite{marcu2023AttachmentInformedDesignDigital}.

\subsection{Ease of Exit: Design for Disengagement}

\Highlight{Chatbots should support an \textbf{ease of exit} because users, especially teens, may struggle to disengage from emotionally immersive interactions.} Our study showed that some teens recognized their own overuse of chatbots (§\ref{sec:RQ3_internal}), while others struggled to disengage despite wanting to (§\ref{sec:RQ2_relapse}). This suggests that current systems do not sufficiently support user agency in disengagement and that design could provide better tools for users to manage and exit these relationships on their own terms. Designers must not only consider how interactions begin and how users can carry out their tasks effectively, but also how they can support an ease of exit. Generative AI makes disengagement harder because it can produce endless content on demand. \Highlight{As our results showed, some teens described the platform’s unlimited possibilities as a reason they stayed engaged for long periods (§\ref{sec:RQ1_creative}), and many reported checking the app constantly throughout the day (§\ref{sec:RQ2_conflict}; §\ref{sec:RQ2_salience}).} The final principle of our framework addresses this issue by proposing ways to support healthier disengagement. \Highlight{Recent platform-level changes highlight this urgency. After our study concluded, Character.AI introduced strict age restrictions for minors \cite{mcmahon2025CharacteraiBanTeens}. This sudden ban affected many teens who were heavily attached to their chatbot companions \cite{dupre2025CharacterAIUsersFull}.}

What separates constructive from harmful disengagement is the degree of agency users retain~\cite{obrien2022RethinkingDisengagementHumancomputer}. Current moderation strategies often rely on rigid restrictions, such as time limits, app locks, or usage reports~\cite{team2024HowCharacteraiPrioritizesa}, and some jurisdictions have experimented with legal controls~\cite{lomas2023ReplikaVirtualFriendship}. Yet such approaches can undermine autonomy and lead to negative experiences. In our study, some users reported abandoning platforms altogether after new censorship measures (§\ref{sec:RQ3_reduced}), pointing to the unintended harms of abrupt, top-down interventions. Sudden restrictions may even cause users to feel that their AI companion has “died” or been replaced, leading to distress or grief~\cite{defreitas2024LessonsAppUpdate}. \Highlight{Our results include posts in which teens reacted to removed or restricted bots with language of loss and grief, describing the characters as \textit{‘gone,’} saying it \textit{‘hurts,’} or feeling they had \textit{‘abandoned’} someone important (§\ref{sec:RQ2_withdrawal}). Public reactions to the Character.AI ban showed a similar pattern. Many teens reported confusion, distress, or withdrawal when access ended suddenly \cite{dupre2025CharacterAIUsersFull}. Experts also warned that abrupt removal can harm teens who relied on these chatbots for daily support \cite{dembosky2025AiSafetyExpert}. These developments support our argument that forced disengagement can trigger shock rather than relief.}

A more effective path is to design disengagement mechanisms that preserve autonomy while taking advantage of the flexibility of language models. \Highlight{Many teens in our dataset expressed a desire to reduce their use voluntarily and sought guidance from others on how to manage or stop their reliance (§\ref{sec:RQ3_internal}; §\ref{sec:RQ2_relapse}). This indicates that designs supporting user-led disengagement may align more closely with teens’ actual needs.} Prior research in video games, for example, showed that players felt less frustration when they were able to pause or save progress before leaving, rather than being forced to quit abruptly mid-quest~\cite{tyack2020MasterclassRethinkingSelfDetermination, davies2016EvaluatingExistingStrategies}. Translating this to chatbot design means that conversations should offer natural closure points, allowing users to complete a thought or goal before disengaging instead of being cut off suddenly and being able to save the state of the conversation. 

Similarly, work on social media use among teens has shown that subtle nudges, delivered in casual and nonjudgmental language, can help users act more intentionally and reduce mindless engagement~\cite{davis2023SupportingTeensIntentional, agha2025SystematicReviewDesignbased}. Tools that help teens grow by managing risks, rather than shielding them through strict control, are more likely to be effective \cite{wisniewski2017ParentalControlVs}. By contrast, systems grounded in surveillance or rigid restriction frequently erode trust and drive disengagement \cite{alelyani2019ExaminingParentChild, ghosh2018SafetyVsSurveillance}, especially since teens value privacy around emotionally sensitive topics \cite{dumaru2025OneSizeDoesnt}. Building on these insights for chatbot design means avoiding blunt notifications or prescriptive commands. Instead, chatbots could weave supportive cues into everyday dialogue that encourage users to reconnect with offline life. During casual exchanges, the system might highlight progress the user has made, suggest opportunities for social contact, or propose light activities outside the platform. Interactions with chatbots could also be designed to support self-reflection by helping users name emotions, revisit meaningful events, and explore their thoughts in their own words, fostering emotional awareness and a stronger sense of self \cite{seo2024ChaChaLeveragingLarge, kim2024MindfulDiaryHarnessingLarge, song2024ExploreSelfFosteringUserdriven}. \Highlight{These strategies match recent recommendations from other child-safety reports that encourage break features, limits on intense interactions, and safeguards that help heavy users reconnect with human support \cite{robb2025TalkTrustTradeoffs}.}

\Highlight{Users in our study also expressed interest in developing healthier habits (§\ref{sec:RQ3_reduced}) and sought support from others (§\ref{sec:RQ3_internal}), indicating openness to tools that promote intentional use. This suggests that beyond conversational prompts}, platforms could integrate usage tracking, emotional check-ins, or personalized limits such as impulse-control tools~\cite{wisniewski2017ParentalControlVs}. \Highlight{Future designers might also consider integrating optional human peer-support features alongside AI interactions. 
In our study, many teens already relied on peers in online communities for advice on quitting or reducing their use (§\ref{sec:RQ3_internal}; §\ref{sec:RQ2_relapse}), suggesting that built-in peer-support systems may align well with existing behaviors. An embedded, age-appropriate peer-support layer, such as anonymous peer groups could provide the human empathy, shared experience, and boundary-setting strategies that AI alone cannot authentically deliver. Such a feature would give teens a way to hear how others navigate similar challenges, feel less isolated in their struggles, and access support grounded in lived experience while maintaining privacy and developmental safety \cite{akter_youthdiscussion_2025, alluhidan_teentalk_2024}. This direction is also consistent with recent guidance that encourages access to human support when AI companions begin to replace real relationships, especially for teens who show signs of unhealthy attachment \cite{robb2025TalkTrustTradeoffs}.} 

Participatory design such as teen advisory councils to create features like customizable moderation settings  \cite{yoon2025ItsGreatBecause,agha2023co,Ekstrand2025co-designing} would further give teens a voice in shaping their digital environment. Instead of sudden content removal, moderation systems could use transitional messages, explanations, or alternatives to help teens process changes. These strategies frame moderation not as a constraint, but as a developmental tool that fosters emotional growth, builds digital resilience, and supports teens’ own approaches to navigating risk.

Unlike gaming or social media, where overuse is often linked to competition, reward cycles, or social validation~\cite{wang2025EmotionalReinforcementMechanism, kaya2024OnlineGamingAddiction}, conversational AI companions provide personalized, reciprocal, and emotionally immersive experiences. Teens in our study did not simply consume or compete; they engaged in ongoing, two-way dialogues with AI characters that could recall past conversations, adapt to their emotional states, and simulate intimacy. These qualities, such as personalization, multimodality, and memory, set AI companions apart from earlier technologies and make overreliance harder to disentangle from authentic-feeling relationships. This underscores the need for further research on the unique characteristics of these relationships and how challenges specific to companion chatbots should be addressed.

%% file: 6_Conclusion.tex
\subsection{Limitations and Future Work}
While our study provides insight into teens’ overreliance on Character.AI, its focus on a single platform limits generalizability to other AI companions with different designs, policies, or user communities. Future research should compare multiple platforms to examine how various features, moderation practices, and interface designs contribute to patterns of overuse. \Highlight{Our dataset consists of voluntarily posted Reddit content, which offers advantages because posts are shared anonymously and at the time the user wanted to share. This reduces social desirability and recall bias compared to surveys or interviews and can lead to more candid reflections~\cite{verbeij2022ExperienceSamplingSelfreports, cinus2025UncoveringSociodemographicFabric}. However, the dataset may overrepresent individuals who recognized their overreliance and chose to share it publicly, while less reflective or less active users may never post. Reddit posts also capture only what users choose to disclose, limiting access to the full context of their everyday experiences~\cite{gaffney2018CaveatEmptorComputational, barikeri2021RedditBiasRealWorldResource}. Future researchers should consider mixed-method approaches, such as surveys or structured interviews, to capture a broader range of perspectives and contex.} \Highlight{As described in our methodology (§\ref{method:data_collection}), our dataset may not include all Reddit posts about chatbot overreliance, yet our analysis reached theoretical saturation for qualitative insights. Future researchers can pursue broader quantitative studies or platform-wide retrieval for exhaustive data coverage.}
 To answer RQ2, we mapped our data to a behavioral addiction model, which may have oversimplified the complex emotional dynamics at play. Future work should compare multiple frameworks, such as attachment~\cite{xie2023FriendMentorLover} and parasocial theory~\cite{horton1956mass}, to test the robustness of different interpretations. The lack of verified demographic data limited our ability to examine how age, gender, or background shape chatbot overreliance, suggesting the need for future studies that directly engage participants and apply best ethical research practices with youth~\cite{razi2024traumainformed}. Lastly, because all authors were fluent in English, we analyzed only English-language posts, which may exclude cultural differences in how teens engage with AI companions and highlights the need for cross-cultural research. While our dataset does not represent teens globally, it reflects a segment of English-speaking adolescents who are active online and engaging with AI companions. This linguistic boundary is common in qualitative internet research and does not invalidate the themes we observed.

\section{Conclusion}

This study shows that teens often begin using AI chatbots for comfort or entertainment, but many gradually become more attached and reliant. Their experiences reflect features of behavioral addiction, with consequences that include disrupted sleep, academic struggles, and strained relationships. Meanwhile, some teens disengage after recognizing harm, reconnecting offline, or encountering platform restrictions. By mapping these patterns to a behavioral addiction framework, the study provides one of the first teen-centered accounts of AI companion overreliance. It highlights the role of emotional attachment and simulated intimacy in shaping teen experiences and introduces the CARE framework as a practical guide for healthier interactions. Looking ahead, future research should compare platforms, use broader methods to capture diverse user experiences, and test design strategies that support constructive disengagement. Recent advances in AI create risks we have never faced before, yet they also open new opportunities. Designers now carry the responsibility to build systems with empathy, nuance, and attention to detail not only to protect teens from harm but also to help them cultivate resilience, growth, and greater fulfillment in their lives.

%% file: appendix.tex
\appendix
\section{Appendix A: Relevancy Coding Prompt}
\label{full_prompt}
We filtered our final dataset of 5,535 posts (prior to relevancy filtering) using OpenAI’s GPT-4o mini to include only posts written by teenagers aged 13 to 17. This was done by automating the relevancy coding process using a Large Language Model (LLM) to determine whether each post was likely authored by a teen.

The following prompt was used to guide the model in performing this classification task:

\begin{quote}
    
Analyze the following post and determine if it was written by a teenager (ages 13--17). Use a strict approach, only allowing posts with very high certainty.

Steps:

1. Clue Identification: Extract explicit clues that directly suggest the writer is 13--17. Examples of clues include references to school, direct self-identifications of age (e.g., "I'm sixteen" or "I'm a minor"), or life experiences typical of this age group. Ignore vague or indirect hints such as slang, dramatics, or informal language.

2. Clue Evaluation: Assign a certainty score (1--5) to each clue based on how strongly it indicates the post is written by a teen. Only consider clues with a score of 4 or 5.

3. Final Decision: If multiple strong clues (score $\geq$ 4) clearly align with teenage behavior, or the post explicitly states their age or suggests they are a teen, respond 'yes.' Otherwise, respond 'no.'

Examples:

Post: "I'm sixteen, and I just started high school. It’s tough but exciting!" (Yes)

Post: "I'm a minor and looking for advice on school-related stress." (Yes)

Post: "I love chatting here, but I wish I had more time between work and family obligations." (No)

Post: "Sometimes I feel overwhelmed by responsibilities, but I manage." (No)

Output:

- Extracted clues.

- Certainty scores for each clue.

- Final decision with a brief justification.

\end{quote}